

\magnification=1200
\global\newcount\meqno
\def\eqn#1#2{\xdef#1{(\secsym\the\meqno)}
\global\advance\meqno by1$$#2\eqno#1$$}
%
\global\newcount\refno
\def\ref#1{\xdef#1{[\the\refno]}
\global\advance\refno by1#1}
\global\refno = 1
\vsize=7.5in
\hsize=5.6in
\tolerance 10000
%
\def\bact{\tilde S_{\beta,k}}

\def\calg{{\cal G}}
\def\calh{ {\cal H}}
\def\calk{ {\cal K}}
\def\calm{ {\cal M}}

\def\calr{ {\cal R}}
\def\ddp{{d^4p\over (2\pi)^4}}
\def\half{{1\over 2}}

\def\mpsi{M_{\psi}}
\def\mvar{M_{\varphi}}
\def\s#1{{\bf#1}}
\def\thru#1{\mathrel{\mathop{#1\!\!\!/}}}
\baselineskip=0.1cm
\baselineskip 12pt plus 1pt minus 1pt
\vskip 2in
\centerline{\bf QUANTUM AND THERMAL FLUCTUATIONS}
\medskip
\centerline{{\bf IN FIELD THEORY}
\footnote{*}{This work is
supported in part by funds
provided by the U. S. Department of Energy (D.O.E.) under contract
\#DE-AC02-76ER03069.}}
\vskip 24pt
\centerline{Sen-Ben Liao,
\footnote{$^\dagger$}{Present address: Department of Physics,
Duke University, Durham, NC\ \ 27708}
Janos Polonyi,
\footnote{$^{\dagger\dagger}$}{Supported in part by the
NSF-PYI grant PHY-8958079.
On leave from CRIP and R. E\"otv\"os University, Budapest, Hungary}
and Dapeng Xu}
\vskip 12pt
\centerline{\it Center for Theoretical Physics}
\centerline{\it Laboratory for Nuclear Science}
\centerline{\it and Department of Physics}
\centerline{\it Massachusetts Institute of Technology}
\centerline{\it Cambridge, Massachusetts\ \ 02139\ \ \ U.S.A.}
\vskip 12pt
\centerline{and}
\vskip 12pt
\centerline{\it Laboratory of Theoretical Physics}
\centerline{\it and Department of Physics}
\centerline{\it Louis Pasteur University}
\centerline{\it 67087\ \ Strasbourg\ \ Cedex\ \ France}
\vskip 1.5in

\vskip 24 pt
\baselineskip 12pt plus 2pt minus 2pt
\centerline{{\bf ABSTRACT}}
\medskip
Blocking transformation is performed in quantum field
theory at finite temperature.
It is found that the manner temperature deforms the renormalized
trajectories can be used to understand better the role played by the quantum
fluctuations. In particular, it is conjectured that domain formation
and mass parameter generation can be observed in theories without spontaneous
symmetry breaking.

\vskip 24pt
\vfill
\noindent CTP\#2143\hfill March 1994
\eject
\medskip
\centerline{\bf I. INTRODUCTION}
\medskip
\nobreak
\xdef\secsym{1.}\global\meqno = 1
\medskip
The understanding of the genuine quantum effects depends crucially on
our ability to describe field fluctuations, the presence of which
attributes to the
fundamental difference between the classical and the quantum systems.
A straightforward method to study fluctuations
for a given system
is by means of introducing some relevant physical parameters which
have influence on the fields. For example, on may consider
a finite region of space-time $\cal R$, and monitor the field fluctuations
within it. If $\phi(x)$ denotes the original field variable which
parameterizes the system, then the exploration of the
physical phenomena in the region $\cal R$ can be facilitated by
introducing the coarse-grained average field:
\eqn\blvi{\phi_{\cal R}={1 \over \Omega_{\cal R}}\int_{\cal R}
d^dx\phi(x),}
where $\Omega_{\cal R}$ is the volume of ${\cal R}$.
The manner in which the quantum fluctuations
affect other degrees of freedom coupled to $\phi(x)$ can then be probed
by studying the $\calr$ dependence of the corresponding distribution
\eqn\distri{\varrho_{\cal R}(\Phi)=<\delta(\Phi-\phi_{\cal R})>
=\int D[\phi]e^{-S[\phi]}\delta(\Phi-\phi_{\cal R}).}

Since it is difficult to compute directly \distri\ in the
conventional perturbation expansion, we adopt an alternative scheme
analogous to the Wilson-Kadanoff blocking procedure in statistical
mechanics \ref\wils.
In this formulation, the averaged blocked field variable in \blvi\ is
written as
\eqn\blfv{\phi_{\cal R}(x)=\int dy\rho_{\cal R}(x-y)\phi(y),}
where $\rho_{\cal R}(x-y)$ is a smearing function which vanishes rapidly
for $x-y\notin{\cal R}$. This leads to a blocked lagrangian
$L_{\cal R}(\phi_{\cal R})$ satisfying
\eqn\bll{<\hat O[\phi_{\cal R}]>
=\int D[\phi]e^{-S[\phi]}{\hat O}[\phi_{\cal R}]
=\int D[\phi_{\cal R}]e^{-S_{\cal R}[\phi_{\cal R}]}{\hat O}[\phi_{\cal R}],}
where
\eqn\bllg{S_{\cal R}=\int d^dx L_{\cal R}.}
In general, $L_{\calr}$ can  be written in the
spirit of Ginsburg-Landau approach as
\eqn\bllglg{L_{\cal R}[\phi]=\sum_{n=0}^\infty U^{(n)}_{\cal R}(\phi(x)),}
where $U^{(n)}_{\cal R}(\phi(x))$ is a homogeneous polynomial of order $n$
in space-time derivatives ${\partial\over\partial x^\mu}$. However, as long
as the inhomogeneities of the configuration $\phi_{\cal R}(x)$ can be
neglected, we then have $U^{(0)}>>U^{(n)}$ for $n\not=0$, and
\distri\ becomes \ref\lp
\eqn\locapr{\varrho_{\cal R}(\Phi)\approx
e^{-\Omega_{\cal R}U_{\cal R}^{(0)}(\Phi)}.}

One notable feature of \distri\ is that it develops non-trivial
structure when $\Omega_{\calr}$ becomes sufficiently small. This can be
realized by examining the `blocked potential' $V_{\cal R}$ in
\eqn\effdis{\varrho_{\calr}(\Phi)=e^{-\Omega_{\cal R}V_{\cal R}(\Phi)}.}
On the one hand, $V_{\calr}$ approaches the usual effective potential
$V_{\rm eff}$ in the thermodynamical limit, $\Omega_{\cal R}\to\infty$, where
all the inhomogeneities of $\phi_{\cal R}$ tend to zero.
On the other hand,
one recovers for $\Omega_{\calr}\to 0$, the bare potential $V_{\rm bare}$
which exhibits stronger divergence than the
kinetic energy term. A negative mass squared term for
sufficiently large values of the UV regulator would imply the existence
of degenerate minima in the potential
$V_{\cal R}(\Phi)$ for small enough $\cal R$.
Such scenario can indeed take place for a symmetric theory in which
the manifest realization
of a global symmetry in the vacuum requires the effective potential to have
a non-degenerate minimum at $\Phi=0$. In this case the distribution is
characterized by a
length scale $\varsigma$ such that $V_{\cal R}(\Phi)$
exhibits a unique minimum at $\Phi=0$ only if the typical size of
$\cal R$ is greater than $\varsigma$ \lp. For smaller domains the potential
$V_{\cal R}(\Phi)$ has minima at $\Phi=\Phi_j\ne 0$ which can be
related to each other by symmetry transformations.

A straightforward explanation of such
behavior of $\varrho_{\calr}(\Phi)$ in \effdis\ can be given by noting
that the theory contains ``domains'' of typical size $\varsigma$
which dominate the path integral. One naturally finds non-vanishing field
average inside each domain, and that
different domains have different symmetry transformations relating
to other field
average values. The smaller the volume $\Omega_{\calr}$ we probe, the
stronger the dominance of the non-vanishing domain values $\Phi$
in the distribution.
We emphasize here that the appearance of degenerate minima in
$V_{\cal R}(\Phi)$ is not the signal of spontaneous breakdown of symmetry.
Contrary to the former case which occurs only for small length scale,
the latter phenomenon refers to the low-energy, large-distance behavior
of the theory in the limit $\Omega_{\cal R}\to\infty$.

In inquiring more details concerning the domain structures in the
symmetric phase, it is also crucial to examine the role of
thermal fluctuations in the system. We shall
choose the region $\cal R$ to be translationally invariant along the
time direction in
Minkowski space-time. The symmetry implies that \blfv\ is an averaged
blocked field
in the three space having a projection of the zero frequency component
of the field variable. The same interpretation applies when going over
to the Euclidean space-time except now $\phi_{\calr}$ contains only
the zero Matsubara frequency modes.
Adopting the latter reference frame, we write
\eqn\blct{\phi_k(\s x)=T\int_0^\beta dy^0\int d^3\s y\rho_k(\s x-\s y)\phi(y),}
where the smearing function is taken to be
\eqn\smfr{\rho_k(\s x)=\int_{|\s p|<k}{d^3\s p\over(2\pi)^3}
e^{i\s p\cdot\s x},}
with $k^{-1}$ being the characteristic linear dimension of
the 3-space volume $\Omega_3$.
Note that \smfr\ which eliminates fast-fluctuating spatial modes
and leaves unconstrained the Matsubara frequency modes is only $O(3)$
symmetric. This can be compared with that of the $O(4)$
invariant smearing function considered in \lp. The full $O(4)$ symmetry can be
recovered only in the limit $k\to 0$ and the effect of temperature is
switched off. Blocking transformation has been used to study various facets of
finite-temperature theory in \ref\wet.

One interesting issue to be addressed here is the competition of the length
scales $k^{-1}$ and $\beta=1/T$. The fact that excitations
are strongly screened at high $T$ leads one to expect
qualitatively the same features for
small $\cal R$ and high $T$ limits, that is,
\effdis\ has similar dependence on the
two length scales. Hence, the value of the most probable field average
increases with $T$. However, this seems to contradict the common conviction
that fluctuations are more "symmetrical" at high
temperature, thereby implying diminishing $\Phi_j$ as $T$ increases.
Indeed, $\Phi_j$ increases as the spatial length
scale, $k^{-1}$ approaches zero.
We shall explicitly show in this paper that the one-loop computation
lends supports the latter scenario.

As the temperature is raised, the contribution of the non-static modes
in Euclidean space-time becomes more suppressed by the large values
of the Matsubara frequencies and a three dimensional effective theory
emerges for the remaining static modes \ref\dimred.
Hence, at high $T$, the distribution \effdis\ reflects the
features of a three dimensional theory.
In the low temperature limit the system is four dimensional since the
$T$ dependence of $\varrho_{\calr}(\Phi)$ is suppressed by a factor
\eqn\supprf{O(e^{-m_L/T}),}
where $m_L$ is the mass gap. The exponent in \supprf\ can be interpreted
as the ratio of
the ``time extent'' and the ``correlation length'' of the system.

Another interesting application one may consider is how the presence of
local domain structure in an overall symmetric background Higgs field
can influence the mass
parameter of a particle propagating through it.
Suppose that a test particle is coupled to the Higgs field $\phi(x)$ and
is moving along $\calr$ which has the shape of a ``tube'' extending along the
time direction with a finite width.
If the potential $V_{\cal R}(\Phi)$ corresponding to
this tube has degenerate minima at the values $\Phi=\Phi_j$, $j=1,\cdots,n$
which are related by the symmetry transformations, then an
observable $\hat O$ associated with the test particle can be
obtained in the mean field
approximation for the Higgs field as
\eqn\meanapr{<<\hat O>>={1\over n}\sum_j<\hat O>_{\phi(x)=\Phi_j},}
where $<\hat O>_{\phi(x)}$ is the expectation value of
$\hat O$ on the background Higgs field $\phi(x)$. In another word,
within the region $\calr$, only the most probable
average values of the Higgs field influences the behavior of the
test particle within the framework of mean field approximation.
The approximation \meanapr\ can be justified by choosing a small $\calr$
such that $\Phi\approx\Phi_j\ne 0$.

Consider now the special case where a scalar test particle
is coupled to the Higgs field in such a way that
its mass squared parameter in the lagrangian, $M^2=\calg\Phi^2$
is the same for all
domains, i.e., $\Phi_j^2=\Phi^2$. The constant $\calg$ is the measure of
the coupling strength between the Higgs field and the test particle.
In this situation one readily finds the mass parameter of the test particle
$M^2\not=0$ even when the Higgs field has
symmetrical vacuum expectation value $<\phi>=0$. In fact, whatever
``important'' Higgs field configuration being taken in \meanapr,
each leads to the same non-vanishing
squared mass parameter. As for the Higgs field, its "blocked" average
decreases as the volume is enlarged to account for the more
independent fluctuations at large separations. This scenario
amounts to the generation of a mass parameter for the test particle
without spontaneous symmetry breaking in the vacuum.

While this simple argument works well for scalars, the situation is
slightly more complicated when the test particle is a fermion.
The reason is that the
coupling between the Higgs field and fermion is linear, contrary to
quadratic coupling for scalars. With
\eqn\ortr{\sum_j\Phi_j=0,}
according to the orthogonality relations for the symmetry group,
the average mass parameter computed by \meanapr\ is vanishing.
For example the usual Higgs field possesses discrete symmetry
$\phi\to-\phi$. Therefore, while the chirality-odd term in the fermion
propagator remain massless, the chirality-even sector in \meanapr\
will be that of a massive fermion.
We again caution the reader that the usual problem
of mass generation is twofold: First, one must give an account
on the appearance of some non-zero mass parameter in the lagrangian
which possesses symmetries characteristic of massless particles. The second
step which can be proceeded with more conventional methods involves
establishing a connection between the mass parameter $m$ and
the actual physical mass $m_{\rm phy}$, i.e.,
the location of the pole in the Minkowski real space-time propagator.
We address only
the first problem in the present work, keeping in mind that
$m_{\rm phy}$ differs from $m$ by additional radiative corrections.

Alas, this type of mass parameter generation mechanism should not be expected
\ref\tsypin\ because the field average is arbitrary small for sufficiently
long observation time. In another words, within a long annular world-tube
$\cal R$, the practically uncorrelated fluctuations of the Higgs field should
yield vanishing most probable average. Nevertheless the mass parameter
might be generated at finite temperature. In fact, at finite temperature
the scalar fields are periodic in the Euclidean time directions
with period length $1/T$. Therefore, the local domains align themselves
and there can be no uncorrelated
fluctuations in the Euclidean time direction at
sufficiently high temperature.
Naturally all one can see in this case is the finite-temperature
Euclidean mass parameter
$m(\Phi_{\beta,k})$ since the real physical mass $m_{\rm phy}(\Phi_{\beta})$
including the thermal contribution can only be determined
by locating the pole of the propagator after an analytic continuation back to
real Minkowski space-time at finite $T$, i.e.,
\eqn\propag{ {1\over {p^2+m^2(\Phi_{\beta,k})}} \longrightarrow
{i\over {p^2-m^2_{\rm phy}(\Phi_{\beta})+i\epsilon}}.}
Although our analyses yield directly a
non-vanishing $m(\Phi_{\beta,k})$, we believe that it
will also contribute to $m_{\rm phy}(\Phi_{\beta})$ in the analytically
continued real-time propagator as well. Notice that in computing $m_{\rm phy}$,
the scale $k$ will be absorbed via a set of self-consistent equations.

The organization of the paper is the following. In Section II we illustrate
the formalism of finite-temperature blocking transformation using the
scalar $\lambda\phi^4$ theory. and generate a renormalization group flow
equation for the finite-temperature blocked potential $U_{\beta,k}$.
The existence of domain structure in the symmetric phase
is examined in the context of Ising model.
The possibility of mass parameter generation without symmetry
breaking is discussed for the Yukawa
model and the scalar QED. Theory with fermionic matter
field coupled to the scalar Higgs potential is presented in Section III where
we
compute the critical scale $k_{cr}$ above which domain structure is detected.
In Section IV where scalar QED is studied, we introduce the blocking scale
$k$ in a gauge invariant manner by inserting a smearing function into
the proper-time
integration variable. Our choice of smearing function which has similar
origin as the Pauli-Villars regularization method, avoids the complication
of UV divergences. We discuss in Section V the issue of dynamical
mass generation
without symmetry breaking by an optimization method in the particle propagator.
It is shown that in the mean field limit, with presence of domain structure
in the background Higgs field, the test particles can acquire a non-vanishing
mass parameter albeit symmetry is preserved. Section VI is reserved for
summary and discussion. In Appendix A, we give an estimate of the width
which defines the effective region traversed by the free test particle. For
completeness, various scenarios for finite-temperature massless scalar QED
are included in Appendix B.

\bigskip
\centerline{\bf II. $\lambda\phi^4$ THEORY}
\medskip
\nobreak
\xdef\secsym{2.}\global\meqno = 1
\medskip
\nobreak
We begin our investigation with the $\lambda\phi^4$ theory described by:
\eqn\lag{{\cal L(\phi)}={1\over2}
(\partial_{\mu}\phi)^2+V(\phi),}
where
\eqn\bpot{V(\phi)={\mu^2\over2}\phi^2
+{\lambda\over4!}\phi^4.}
Before computing the finite temperature blocked action $\bact$,
which is the effective action at the scale $k$ and temperature $T$, we
give a brief review of blocking transformation at zero temperature in
$d=4$ Euclidean space-time.

The first step is to separate the slowly-varying background fields $\sigma$
from the fast-fluctuating modes $\xi$, i.e.,
\eqn\field{\phi(p)=\cases{\sigma(p),&$0 \le p \le k$ \cr
\cr
\xi(p), &$k < p < \Lambda$.  \cr }}
This can be achieved most easily by applying an $O(4)$ symmetric
sharp momentum smearing
function $\rho_k(p)=\Theta(k-p)$.
Given a set of blocked variables $\phi_k(x)$,
the blocked action $\tilde S_k$ can be derived from
\eqn\cactt{ e^{-\tilde S_k[\Phi]}=\int D[\phi]\prod_x
\delta(\phi_k(x)-\Phi(x))e^{-S[\phi]},}
where the field average $\Phi$ of a given block is chosen to coincide with
the slowly varying background since $\phi_k(p)=\rho_k(p)\phi(p)=\sigma(p)$.
One then integrates out the
fast-fluctuating modes $\xi(x)$ using the loop expansion
to obtain $\tilde S_k[\Phi]$:
\eqn\efact{\eqalign{\tilde S_k[\Phi]&=-{\rm ln}\int D[\sigma]
D[\xi]\prod_x\delta(\phi_k(x)-\Phi(x))~{\rm exp}\Bigl\{-S[\sigma+\xi]\Bigr\}\cr
&
=-{\rm ln}\int D[\sigma]\prod_p\delta(\sigma(p)-\Phi(p))\int D[\xi]~{\rm exp}
\Bigl\{-S[\sigma]-{1\over 2}\int_p^{'}\xi(p)K(\sigma)\xi(-p)+\cdots\Bigr\} \cr
&
=-{\rm ln}\int D[\sigma]\prod_p\delta(\sigma(p)-\Phi(p))~{\rm exp}
\Bigl\{-S[\sigma]-{1\over 2}{\rm Tr'}{\rm ln}K(\sigma)+\cdots\Bigr\} \cr
&
=S[\Phi]+~\half{\rm Tr'}{\rm ln}K(\Phi) ,}}
where
\eqn\kern{ K(\Phi) = {{\partial^2 {\cal L}}
\over \partial\phi^2(x)}\Big\vert_{\Phi}=-\partial^2 +V''(\Phi),}
\eqn\spi{ \int_p^{'}=\int_k^{\Lambda}\ddp ,}
and ${\rm Tr'}$
implies that the trace in momentum space is to be carried out for
$k \le p \le \Lambda$, i.e., the modes which are to be eliminated by
blocking transformation. $\tilde S_k[\Phi]$ can readily be seen to
interpolate smoothly between the bare action $S[\Phi]$
and the renormalized one-loop effective action as $k$ evolves from
$\Lambda$ to $0$.

Going to the finite-temperature formalism,
the field periodic in the time with period $\beta$ can be
written as
\eqn\ppx{\eqalign{\phi(x)&={1\over\beta}\sum_{n=-\infty}^\infty\int
{d^3\s p\over(2\pi)^3}e^{-i(\omega_n\tau-\s p\cdot\s x)}\phi(\omega_n,\s p)\cr
&\longrightarrow{1\over\beta}\sum_{n=-\infty}^{\infty}\int
{d^3\s p\over(2\pi)^3}e^{ipx}\phi(p),}}
where $p^{\mu}=(i\omega_n,\s p)$ and
$\omega_n={2\pi n\over\beta},$ the Matsubara frequencies for bosons.
The analogous coarse-grained blocked field is defined as:
\eqn\blctk{\phi_k(\s x)=T\int_0^\beta dy^0\int d^3\s y
\rho_k(\s x-\s y)\phi(y),}
where
\eqn\smfrk{\rho_k(\s x)=\int_{|\s p|<k}{d^3\s p\over(2\pi)^3}
e^{i\s p\cdot\s x}.}
A comparison between the manner in which momentum modes are eliminated
using the $O(4)$ and $O(3)$
symmetric smearing functions is depicted in Fig. 1.
The finite-temperature blocked action $\tilde S_{\beta,k}[\Phi]$,
can be computed in the same manner as the zero temperature case,
with the exception of evaluating the kernel $K(\Phi)$. Summing over
the discrete Matsubara frequency modes, we have
\eqn\uc{\eqalign{
&{1\over2\Omega}{\rm Tr'}{\rm ln}K(\Phi)=
{1\over2\beta}\sum_n\int^{\infty}_k {d^3\s p\over(2\pi)^3}{\rm ln}
\Bigl[\omega_n^2+p^2+V''(\Phi)\Bigr]\cr
&={1\over 2\beta}\int^{\infty}_k{d^3\s p\over (2\pi)^3}
\Biggl\{\beta\sqrt{p^2+V''(\Phi)}+2{\rm ln}
\Bigl[1-e^{-\beta\sqrt{p^2+V''(\Phi)}}\Bigr]\Biggr\}.}}
By separating the zero temperature contribution in \uc\ via the relation
\eqn\mkf{\int^{\infty}_{-\infty}{dp_0\over 2\pi}
{}~{\rm ln}\bigl[ p_0^2+E^2\bigr]=E,}
which holds up to an infinite $E$-independent constant, one obtains
\eqn\ftsep{\eqalign{U_{\beta,k}(\Phi)&=V(\Phi)
+{1\over2}\int_k^{\infty}{d^3\s p\over(2\pi)^3}\int_{-\infty}^\infty
{dp^0\over2\pi}{\rm ln}\Bigl[p^2+V''(\Phi)\Bigr]\cr
&+T\int_k^{\infty}{d^3\s p\over(2\pi)^3}{\rm ln}
\Bigl[1-e^{-\beta\sqrt{p^2+V''(\Phi)}}\Bigr].}}
This shows explicitly how the blocked potential for the
static Euclidean modes is given by the analytic continuation
of the blocked potential for the static modes in Minkowski
space-time. In fact, the latter contains the integration over
each frequency and is just the second term in the right hand side of \ftsep.

Since the manipulation above applies to arbitrary potential
$V(\Phi)$, the renormalization group equation for the finite-temperature local
potential $U_{\beta,k}(\Phi)$ can be obtained by differentiating
\ftsep\ with respect to $k$:
\eqn\rgft{k{{\partial U_{\beta,k}}\over {\partial k}}
=-{k^3\over4\pi^2}\sqrt{k^2+U''_{\beta,k}}
-T{k^3\over2\pi^2}{\rm ln}\Bigl[1-e^{-\beta\sqrt{k^2+U''_{\beta,k}}}\Bigr].}
The higher loop contributions are vanishing in the renormalization group
equation \rgft\ since they are suppressed by ${{\partial k}\over k}\to0$.
To obtain a flow equation for the full theory, however, one needs to
take into accounts the wave function renormalization
constant as well as the higher order derivative terms
in the blocked lagrangian. The simplifications, we believe, represent
only small errors in the infrared limit of the four dimensional theories.

While at low temperature, $T<<\sqrt{k^2+U''_{\beta,k}}$, \rgft\ reduces to
\eqn\rgftl{k{{\partial U_{\beta,k}}\over {\partial k}}=
-{k^3\over4\pi^2}\sqrt{k^2+U''_{\beta,k}},}
in the high $T$ limit where $T>>\sqrt{k^2+U''_{\beta,k}}$, we find
\eqn\rgfth{k{{\partial U_{\beta,k}(\Phi)}\over {\partial k}}=
-T{k^3\over4\pi^2}{\rm ln}\Bigl[{{k^2+U''_{\beta,k}(\Phi)}\over
{k^2+U''_{\beta,k}(0)}}\Bigr].}

These expressions could have been guessed easily. At high temperature
the large Matsubara frequency suppresses the contribution of the
non-static modes. This can be see in the first equation of \uc\
where the terms with $n\not=0$ have diminishing dependence on $\Phi$.
When the contribution of the non-static modes is
altogether neglected then
the static modes are described by a three dimensional
theory with the potential $\tilde U_k=\beta U_{\beta,k}$.
The natural field variable
is $\tilde\Phi=\sqrt{\beta}\Phi$ in three dimensions and the
renormalization group equation of the three-dimensional theory becomes
(c.f. \rgfth\ ):
\eqn\thrrg{k{\partial{\tilde U_k(\tilde\Phi)}\over{\partial k}}
=-{k^3\over4\pi^2}{\rm ln}
\Bigl[{{k^2+{\ddot {\tilde U_k}}(\Phi) }\over
{k^2+{\ddot{\tilde U_k}}(0) }}\Bigr],}
where the dot denotes differentiation with respect to $\tilde\Phi$.
By writing the renormalization group equation \rgft\ as
\eqn\rgfts{\eqalign{k{{\partial U_{\beta,k}}\over {\partial k}}
&=-{k^3\over4\pi^2}\sqrt{k^2+U''_{\beta,k}}\cr
&-T{k^3\over4\pi^2}\Biggl\{2{\rm ln}\Bigl[1-e^{-\beta\sqrt{k^2+U''_{\beta,k}}}
\Bigr]-{\rm ln}\Bigl[{{k^2+U''_{\beta,k}}\over
{k^2+U''_k(0)}}\Bigr]\Biggr\}\cr
&-T{k^3\over4\pi^2}{\rm ln}\Bigl[{{k^2+U''_{\beta,k}}\over
{k^2+U''_k(0)}}\Bigr],}}
we separate the $T=0$, the finite temperature and the $T=\infty$ contributions.
While first two lines on the right-hand side correspond to
the non-static modes, the last term
contains the contribution of the static modes to the renormalization of the
potential. It is rather surprising to find among these terms
a non-logarithmic temperature-independent contribution. The high
temperature system is close to the three dimensional one and modifications
in the infrared region becomes more enhanced for lower dimensions.
Such qualitative difference
is clearly visible between the first and the last line since the
power-like function increases more rapidly than that of logarithm.

Instead of integrating out the renormalization group equation \rgft,
we can compute the blocked potential easily in the independent-mode
approximation which yields:
\eqn\upk{\eqalign{U_{\beta,k}(\Phi)&={\mu_R^2\over2}\Phi^2
\Bigl(1-{\lambda_R\over64\pi^2}\Bigr)
+{\lambda_R\over4!}\Phi^4\Bigl(1-{9\lambda_R\over64\pi^2}\Bigr)\cr
&+{1\over32\pi^2}\Biggl\{-k\Bigl(2k^2+\mu_R^2+{1\over2}\lambda_R\Phi^2\Bigr)
\Bigl({k^2+\mu_R^2+{1\over2}\lambda_R\Phi^2}\Bigr)^{1/2}\cr
&+\Bigl(\mu_R^2+{1\over2}\lambda_R\Phi^2\Bigr)^2{\rm ln}\Bigl[{k+{\sqrt
{k^2+\mu_R^2+\lambda_R\Phi^2/2}}\over \mu_R}\Bigr]\Biggr\}\cr
&+{1\over2\pi^2\beta}\int_k^{\infty}dpp^2{\rm ln}
\Bigl[1-e^{-\beta\sqrt{p^2+\mu_R^2+\lambda_R\Phi^2/2}}\Bigr],}}
where $\mu_R^2$ and $\lambda_R$ are the renormalized parameters
satisfying
\eqn\hemo{\cases{\eqalign{\mu^2_R&=
{\partial^2U_{\beta,k}\over\partial\Phi^2}\Big\vert_{\Phi=1/{\beta}=k=0}\cr
\lambda_R&={\partial^4U_{\beta,k}\over\partial\Phi^4}
\Big\vert_{\Phi=1/{\beta}=k=0.} \cr}}}
One can however write down the general effective scale- and
temperature-dependent
mass squared $\mu^2_R(\beta,k)$ and coupling constant $\lambda_R(\beta,k)$:
\eqn\effmass{\eqalign{ \mu_R^2(\beta,k)&=\mu_R^2-{\lambda_R\over
{64\pi^2\sqrt{k^2+\mu_R^2}}}\Biggl\{4k^3+\mu_R^2\Bigl(3k-\sqrt{k^2+\mu_R^2}
\Bigr)-{\mu_R^4\over {k+\sqrt{k^2+\mu_R^2}}} \cr
&
-4\mu_R^2\sqrt{k^2+\mu_R^2}
{\rm ln}\Bigl({{k+\sqrt{k^2+\mu_R^2}}\over \mu_R}\Bigr)\Biggr\}
+{\lambda_R\over 4\pi^2\beta^2}\int_{\beta\sqrt{k^2+\mu_R^2}}^{\infty}
{}~dx{{\sqrt{x^2-\beta^2\mu_R^2}}\over {e^x-1}} ,}}
and
\eqn\effcou{\eqalign{ \lambda_R(\beta,k)&=\lambda_R-{3\lambda_R^2\over
16\pi^2}\Biggl\{{{k\bigl[2k+(2k^2+\mu_R^2)(k^2+\mu_R^2)^{-1/2}\bigr]}\over
{(k^2+\sqrt{k^2+\mu_R^2})^2}}-{\rm ln}\Bigl({{k+\sqrt{k^2+\mu_R^2}}\over
\mu_R}\Bigr)\Biggr\} \cr
&
-{3\lambda_R^3\over 8\pi^2}\int_{\beta\sqrt{k^2+\mu_R^2}}^{\infty}
{}~dx{{\sqrt{x^2-\beta^2\mu_R^2}}\over x^2}~{{e^x-1+xe^x}\over {e^x-1}} ,}}
where $x=\beta\sqrt{p^2+\mu_R^2}$.
In the limit $1/{\beta}=k=0$, we readily recover the renormalized
quantities and the Coleman-Weinberg potential \ref\col:
\eqn\cop{\eqalign{U_{\rm eff}(\Phi)
&={\mu_R^2\over2}\Phi^2\Bigl(1-{\lambda_R\over64\pi^2}\Bigr)
+{\lambda_R\over4!}\Phi^4\Bigl(1-{9\lambda_R\over64\pi^2}\Bigr)\cr
&+{1\over64\pi^2}\Bigl(\mu_R^2+{1\over2}\lambda_R\Phi^2\Bigr)^2{\rm ln}
\Bigl({\mu_R^2+\lambda_R\Phi^2\over\mu_R^2}\Bigr).}}
If one is only interested in the small $\beta$ and large $k$ physics, it is
possible to approximate
the above integral expressions. Setting $k^2\gg\mu_R^2
+{1\over2}\lambda_R\Phi^2$, we find, keeping only the leading order
contributions:
\eqn\ugk{U_{\beta,k}(\Phi) \approx{1\over2}\mu_R^2(\beta,k)\Phi^2
+{\lambda_R \over 4!}\Phi^4 ,}
\eqn\skm{\mu_R^2(\beta,k)= \mu_R^2
-{\lambda_R \over {16\pi^2}}k^2
+{{\lambda_R} \over 4\pi^2\beta^2}\sum_{n=1}^{\infty} e^{-n\beta k}
\Bigl({1\over n^2}+{{\beta k}\over n}\Bigr)+\cdots,}
where the following integrations have been used \ref\arfken:
\eqn\debyef{ \int_0^r{du~u^{\ell}\over {e^u-1}}=r^{\ell}\Bigl[{1\over \ell}
-{r\over 2(\ell+1)}+\sum_{n=1}^{\infty}{{B_{2n}r^{2n}}\over
{(2n+\ell)(2n)!}}\Bigr] \qquad (\ell \ge 1),}
\eqn\debyeff{ \int_r^{\infty}{du~u^{\ell}\over {e^u-1}}=\sum_{n=1}^{\infty}
e^{-n r}\Bigl[{r^{\ell}\over n}+{{\ell r^{\ell-1}}\over \ell^2}
+{{\ell(\ell-1)r^{\ell-2}}
\over n^3}+\cdots +{{\ell !}\over n^{\ell+1}}\Bigr] \qquad (\ell\ge 1).}
Note that the diverging structures in $\mu_R^2(\beta^{-1}=0,k)$
and $\lambda_R(\beta^{-1}=0,k)$ resemble that of the bare quantities.
However, discrepancy exists between the coefficients since for
renormalizing the theory, it suffices to use $T$-independent counterterms
which are usually derived with an $O(4)$ symmetric smearing function
$\rho_k(x)$. However, $\rho_k(x)$ is only $O(3)$ symmetric for the
present theory.

The most probable value of the average field according to \locapr\ satisfies
\eqn\mit{ 0 = {{\partial U_{\beta,k}}\over {\partial \Phi}}
= \Phi \Bigl[ \mu_R^2(\beta,k) + {\lambda_R \over 6}\Phi^2 \Bigr],}
which shows that in addition to $\Phi=0$ , there exists other non-trivial
solutions if $\mu_R^2(\beta,k) < 0$.
Consider first the case $\mu_R^2 >0$ such that the theory is in the
symmetric phase. Then $U_{\beta, k}(\Phi)$
has only a unique minimum at $\Phi=0$ for positive $\mu_R^2(\beta,k).$
However, for a fixed $\beta$, there always exists a critical value
$k_{cr} \sim O(\mu_R/{\sqrt \lambda_R})$
beyond which $\mu_R^2(\beta,k)$ becomes negative. When this happens,
$U_{\beta,k}(\Phi)$ has non-trivial minima at
\eqn\lmi{\pm\Phi_{\beta,k}\approx\pm\Biggl(-{6\mu_R^2(\beta,k)\over\lambda_R}
\Biggr)^{1\over 2}.}
This can be interpreted as having non-vanishing average coarse-grained
background fields that are distributed primarily
around ${+ \Phi_{\beta,k}}$ or ${-\Phi_{\beta,k}}$ with equal
probability, giving an overall zero average field ${\Phi=0}.$
In the same time the system as a whole retains its full symmetry.

Do we really have such a peculiar distribution of the collective variable
$\Phi_{\beta,k}$ in the path integral for the Higgs field?
To understand this issue better, let us turn to the four dimensional
Ising model in the high temperature phase. The distribution of the
average magnetization $s_v$ for the lattice of volume $v$ is given by
\eqn\avmis{\varrho_v(\Phi)=<\delta(0;v(\Phi-s_v))>
=\sum_{\{s_a\}}\delta(0;v(\Phi-s_v))e^{-\beta H},}
where $H$ is the hamiltonian and $\delta(i;j)$ is the Kronecker delta and
\eqn\avs{s_v={1\over v}\sum_{a\in v}s_a.}
In the ultraviolet limit with $v=1$, it is easy to find individual spins
according to \avmis:
\eqn\iso{\varrho_1(\Phi)={1\over2}\bigl(\delta(1;\Phi)+\delta(-1;\Phi)\bigr).}
On the other hand, in the infrared limit where $v\to\infty$,
\avmis\ gives the distribution
of the global magnetization which, in the symmetrical phase,
has a maximum at $\Phi=0$ and a curvature $m_L^2v$, where $m_L$ is the
inverse correlation length, i.e. mass gap in lattice spacing units.
It is not difficult to understand such "transmutation" of distribution
from the double-peak to the single-peak shape as we
move towards the infrared domain. In fact, if we divide the space-time into
hypercubes with characteristic linear width $k^{-1}$, then for
$k^{-1}>m_L^{-1}$ the magnetization
$s_{k^{-4}}$ can be obtained by averaging over the magnetization
$s_{m_L^{-4}}$ of hypercubes with size $m_L^{-4}$ which are inside of the
hypercube of size $k^{-4}$. The correlation of the magnetization
of these smaller hypercubes is suppressed by $O(e^{-m_L/k})$ and the
central limit theorem can be invoked to give an approximate
description of the
distribution of their average, $s_{k^{-4}}$. Thus, $\varrho_k(\Phi)$
will be a Gaussian distribution centered at $<s>=0$. Cross-over between
the double-peak feature in the ultraviolet and the single-peak shape in the
infrared limit can take place at
\eqn\cross{k_{cr}=O(m_L).}
The long-range instability present in the symmetry broken phase
of the theory would, however, prevents the distribution function
$\varrho(\Phi)$ from approaching the Gaussian limit in the infrared regime.
We give a plot of $U_{\beta,k}(\Phi)$ in Fig. 2 for various values of $k$
at a fixed $T$.

In the above treatment, $k_{cr}$ is qualitatively the scale in which the
``effective mass'' becomes zero. We associate it with the
``correlation length'' $\varsigma$
which defines the domain size via $\varsigma\sim k_{cr}^{-1}$. Note that
this definition of correlation length differs from the usual one.
Clearly, $\varsigma$ is $T$-dependent. This implies that
$k_{cr}$ also depends on $T$. The manner in which $\varsigma$ varies
with $T$ can be seen from:
\eqn\ppvar{ T{{\partial {\rm ln}\varsigma}\over {\partial T}}
= -{2\over {{\rm coth}\bigl({\nu\over 2}\bigr)}}\Bigl[
{1\over {e^{\nu}-1}}
+{2\over \nu}\sum_{n=1}^{\infty}{e^{-n\nu}\over n}\Bigl(
1+{1\over {n\nu}}\Bigr)\Bigr],}
where $\nu=\beta\varsigma^{-1}$. The negative sign in \ppvar\
shows that domain size decreases with increasing $T$.
In Fig. 3, we demonstrate that for a large $k$ where the blocked potential
is double-welled, it is always possible to adjust $T$ such that
for $T > T_k$, $U_{\beta,k}(\Phi)$ again has unique minimum at $\Phi=0$.
Physically, this implies that an increase
of temperature leads to a greater extent of randomness and results
in a smaller domain size. That the two scale parameters, $k$ and $T$
generate opposite kind of changes in the distribution function can
easily be seen from the way $\Phi_{\beta,k}$ varies. Namely,
using \skm, one has:
\eqn\pvar{ k{{\partial \Phi^2_{\beta,k}}\over {\partial k}}
=-\Bigl({6\over\lambda_R}\Bigr)~k{{\partial\mu_R^2(\beta,k)}\over
{\partial k}}
={3k^2\over 4\pi^2}\Bigl[1+2\sum_{n=1}^{\infty}e^{-n\vartheta}\Bigr]
={3k^2\over 4\pi^2}~{\rm coth}\Bigl({\vartheta\over 2}\Bigr) > 0,}
and
\eqn\pvarr{ T{{\partial \Phi^2_{\beta,k}}\over {\partial T}}
=-\Bigl({6\over\lambda_R}\Bigr)~T{{\partial\mu_R^2(\beta,k)}\over
{\partial T}}
=-{3k^2\over 2\pi^2}\Bigl[{1\over {e^{\vartheta}-1}}
+{2\over \vartheta}\sum_{n=1}^{\infty}{e^{-n\vartheta}\over n}\Bigl(
1+{1\over {n\vartheta}}\Bigr)\Bigr] < 0,}
where $\vartheta=\beta k$. Adding these two terms together, one finds:
\eqn\pva{ R= k{{\partial \Phi^2_{\beta,k}}\over {\partial k}}
+T{{\partial \Phi^2_{\beta,k}}\over {\partial T}}
={3k^2\over 4\pi^2}\Bigl[1-{4\over \vartheta}\sum_{n=1}^{\infty}
{e^{-n\vartheta}\over n}\Bigl(1+{1\over \vartheta}\Bigr)\Bigr],}
which has a maximum value of $R={3k^2\over 4\pi^2}$ at
$\vartheta=\infty ~(T=0)$, and decreases
as $T$ is raised. Numerically, the critical value at which $R=0$
is found to be $\vartheta_{cr}=1.53$. Below $\vartheta_{cr}$,
$R$ becomes negative.
This shows that with the given initial values for $T$ and $k$ which give
a non-vanishing $\Phi_{\beta,k}$, an increase in $T$ that shifts
$\Phi_{\beta,k}$ toward the origin can be overcome with a corresponding
increase in $k$ at a ratio larger than or equal to $\vartheta_{cr}$.

While \effmass\ includes the leading order self-energy for the particle
in the heat bath, one must be careful in interpreting
the temperature effect on the coupling constant using
\effcou. Naive differentiation of \effcou\ would give:
\eqn\vart{ T{{\partial\lambda_R(\beta,k)}\over {\partial T}}
=-{3\lambda_R^2\over 8\pi^2}\int_{\beta\sqrt{k^2+\mu_R^2}}^{\infty}
{}~dx\sqrt{x^2-\beta^2\mu_R^2}~{{e^x(e^x+1)}\over {(e^x-1)^3}},}
which shows that within the framework of one-loop approximation,
$\lambda_R(\beta,k)$ dimishes with increasing $T$. This feature can be
substantiated by the physical observation of screening effect.
However, it is not clear how $\lambda_R(\beta,k)$ behaves at very high $T$.
As pointed out in \ref\wein\ and \ref\pen,
the validity of perturbation theory becomes questionable at very high
temperature, and that higher loop effects may also become important
\ref\funa.

In passing, we show how the above arguments are connected to
the case where discrete reflection symmetry $\Phi
\rightarrow -\Phi$ is spontaneously broken by requiring $\mu_R^2 <0$. In
this symmetry broken regime, the
potential $U_{\rm eff}(\Phi)$ in \cop\ develops imaginary part
for small $\Phi$ due to the
negative argument $\mu_R^2+\lambda_R\Phi^2/2$ in the logarithm.
This instability indicates that the dominant field configurations are
not homogeneous anymore. The vacuum state is then dominated by configurations
where the sign of the field is different.
However, if we consider a blocked system instead,
then the imaginary part of the local
potential can be "smeared out" by choosing a scale
$k^2 > \mu_R^2 + \lambda_R\Phi^2/2.$ For such $k$ the domain
size is smaller than the inhomogeneous fluctuations of the vacuum and
the instability does not show up in the blocked potential.
In this symmetry-broken regime where
$\mu_R^2(\beta,k) <0$ at low temperature, $U_{\beta,k}(\Phi)$ has
non-trivial local minima at $\pm\Phi_{\beta,k}.$ Now by universality
we anticipate a second order phase transition at high temperature
which restores the symmetry. The critical temperature above which
symmetry is restored is given by
\eqn\syr{ \mu_R^2(\beta_c,k=0) = 0 ,}
which yields \ref\jack\
\eqn\syt{ {1\over\beta_c^2}\approx -{24\mu_R^2\over\lambda_R}.}
Indeed $\beta_c$ is a physical quantity independent of $k$. Above
the critical temperature, the potential again becomes
reflection-symmetric with a
minimum at $\Phi=0.$ Nevertheless, if we continue to increase $k$,
field fluctuations around some non-trivial minima
will eventually occur.

\bigskip
\nobreak
\medskip
\centerline{\bf III. YUKAWA MODEL}
\xdef\secsym{3.}\global\meqno = 1
\medskip
\medskip
We shall now consider the Yukawa model with massless fermions coupled to the
$\lambda\phi^4$ scalar sector. The lagrangian for the theory is
\eqn\llag{ {\cal L}(\bar\psi,\psi,\phi)={\bar\psi}\calm(\phi)\psi+ V(\phi),}
where
\eqn\matr{ \calm(\phi)={\thru {\partial}}+ g\phi,}
\eqn\scal{V(\phi)= {1 \over 2}(\partial_{\mu} \phi)^2
+ {\mu^2\over 2} \phi^2 + {\lambda\over 4! } \phi^4 ,}
and $g$ is the Yukawa coupling constant.
Following the presentation of Section II, we decompose
$\phi$ into the fast modes $\xi$ and the slow background $\sigma$, and
expand the fermionic fields about their classical configurations $\bar\eta$
and $\eta$. The lagrangian can then be rewritten as:
\eqn\tre{\eqalign{{\cal L}(\bar\eta+\bar\zeta,\eta+\zeta,\sigma+\xi) &
={\cal L}(\bar\eta,\eta,\sigma) +\bar\zeta\calm(\sigma)\zeta
+ g\xi\Bigl(\bar\eta\zeta+\bar\zeta\eta\Bigr)
+{1 \over 2}\xi\calk(\sigma)\xi \cr
&
={\cal L}(\bar\eta,\eta,\sigma)+\bar\zeta'\calm(\sigma)\zeta'
+{1\over 2}\xi\calk'(\bar\eta,\eta,\sigma)\xi ,}}
where
\eqn\dert{\calk'(\bar\eta,\eta,\sigma)=-\partial^2
+V''(\sigma)+2g^2\bar\eta({\thru\partial}
+ g\sigma)^{-1}\eta=\calk(\sigma)+2g^2\bar\eta({\thru\partial}
+ g\sigma)^{-1}\eta,}
and
\eqn\zdd{\cases{\eqalign{ \bar\zeta' &= \bar\zeta
+ g\xi\bar\eta\calm(\sigma)^{-1} \cr
\zeta' &=\zeta+g\xi\calm(\sigma)^{-1}\eta . \cr}}}
The equations above show that the effect of coupling between the
scalar and the fermion sectors is
to add to the original scalar field determinant $\calk$ an extra term
proportional to the background fermion fields.

With the Jacobian of such field transformation being unity:
\eqn\jaco{ D[\zeta']D[\bar\zeta']=D[\zeta]D[\bar\zeta],}
one can
easily perform the Gaussian integrations and obtain:
\eqn\efact{\eqalign{\tilde S_k[\bar\eta,\eta,\Phi]&=-{\rm ln}\Biggl\{
\int D[\sigma]D[\xi] D[\zeta']D[\bar\zeta']
\prod_p\delta(\sigma-\Phi) \cr
&
\times
{}~{\rm exp}\Bigl(
-S[\bar\eta,\eta,\sigma]-\int_p\bar\zeta'\calm(\sigma)\zeta'
-{1\over 2}\int_p^{'}\xi\calk'(\bar\eta,\eta,\sigma)\xi+\cdots\Bigr)
\Biggr\} \cr
&
=S[\bar\eta,\eta,\Phi]-{\rm Tr}{\rm ln}\calm(\Phi)
+\half{\rm Tr'}{\rm ln}\calk'(\bar\eta,\eta,\Phi).}}
Note that the one-loop contributions from the fermion and the scalar fields
appear opposite in sign.

Concentrating only on the scalar sector, we set the
background fermion fields to be zero and obtain:
\eqn\ewbd{\eqalign{U_{\beta,k}(\Phi)&={1\over2}\Bigl(\mu^2_R+\delta\mu^2\Bigr)
\Phi^2+{1\over4!}\Bigl(\lambda_R+\delta\lambda\Bigr)\Phi^4 \cr
&
+ {1\over 2\beta}\sum_n \int_k^{\infty} {dp p^2\over 2\pi^2} {\rm ln}
\Bigl[\omega_n^2+p^2 + \mu_R^2 + {\lambda_R \over 2}\Phi^2\Bigr] \cr
&
- {2\over \beta}\sum_n \int_0^{\infty} {dp p^2\over 2\pi^2} {\rm ln}
\Bigl[\omega_n^2+p^2 + g^2\Phi^2 \Bigr] . }}
Imposing the usual off-shell renormalization condition
\eqn\sig{\lambda_R={{\partial^4 U_k}\over{\partial\Phi^4} }
\Big\vert_{\Phi=\mu_R}}
then leads to
\eqn\pke{\eqalign{U_k(\Phi)&={\mu_R^2\over 2}\Phi^2\Bigl(1-{\lambda_R\over
64\pi^2} \Bigr)+{\lambda_R\over 4!}\Phi^4\Bigl(1-{9\lambda_R\over
64\pi^2}\Bigr)
-{g^4\Phi^4\over 16\pi^2}\Bigl[
{\rm ln}\Bigl({\Phi^2\over\mu_R^2}\Bigr)-{25\over 6} \Bigr] \cr
&
+{1\over 32\pi^2}\Biggl\{ -k\Bigl(2k^2+\mu_R^2+{\lambda_R \over 2}\Phi^2\Bigr)
\Bigl({ k^2 + \mu_R^2+{\lambda_R \over 2}\Phi^2}\Bigr)^{1/2} \cr
&
+\Bigl(\mu_R^2+{\lambda_R \over 2}\Phi^2\Bigr)^2
{\rm ln}\Bigl[{ { k+ \sqrt{ k^2+\mu_R^2+\lambda_R\Phi^2/2}}
\over \mu_R}\Bigr]\Biggr\} \cr
&
+{1 \over {2\pi^2\beta}}\int_k^{\infty} dp p^2 {\rm ln}
\Bigl[1- e^{- {\beta \sqrt { p^2+\mu_R^2+\lambda_R\Phi^2/2 }}}\Bigr] \cr
&
- {2 \over {\pi^2\beta}} \int_0^{\infty} dp p^2 {\rm ln}
\Bigl[1- e^{- {\beta \sqrt {p^2+g^2\Phi^2 }}}\Bigr]   .}}
Keeping only the leading terms in the limit where
$k^2,T^2 \gg \mu_R^2+\lambda_R\Phi^4/2$
and $g^2\Phi^2,$ we arrive at
\eqn\ubkk{U_{\beta,k}(\Phi) \approx{1\over2}\mu_R^2(\beta,k)\Phi^2
+{\lambda_R \over 4!}\Phi^4 -
{g^4\over 16\pi^2}\Phi^4\Bigl[{\rm ln}\Bigl({\Phi^2\over\mu_R^2}\Bigr) -
{ 25 \over 6} \Bigr] ,}
where, by the help of \debyeff,
\eqn\tbm{\mu_R^2(\beta,k)= \mu_R^2 - {\lambda_R \over {16\pi^2}}k^2
+{1 \over 12\beta^2} \Biggl[ {3\lambda_R\over \pi^2}\sum_{n=1}^{\infty}
e^{-n\beta k}\Bigl({1\over n^2}+{{\beta k}\over n}\Bigr)
-4g^2 \Biggr]. }
Here, one must be careful in choosing $\lambda_R$ and $g$, to avoid
$U_{\beta,k}(\Phi)$ from becoming unbound from below \ref\lin. However,
the general feature of field fluctuations around non-trivial
minima at some observational length scale $k^{-1}$ remains unaffected,
since the field where the local maximum of $U_{\beta,k}(\Phi)$ is
attained, lies considerably far away from the origin.
The condition for existence of non-trivial minimum for the potential
implies
\eqn\exis{ 0 = \mu_R^2(\beta,k) + \Phi^2\Bigl( {\lambda_R \over 6} +
{11g^4 \over 12\pi^2} \Bigr) - {g^4\Phi^2 \over 4\pi^2}{\rm ln}
\Bigl({\Phi^2 \over \mu_R^2}\Bigr) ,}
which becomes, for $k=0$,
\eqn\exiss{ 0 = \mu_R^2 +{{\lambda_R - 8g^2} \over 24\beta^2}
+ \Phi^2\Bigl( {\lambda_R \over 6} +
{11g^4 \over 12\pi^2} \Bigr) - {g^4\Phi^2 \over 4\pi^2}{\rm ln}
\Bigl({\Phi\over \mu_R^2}\Bigr) .}
We shall choose $\lambda_R > 8g^2$ such that an increase in temperature will
tend to restore the symmetry. We illustrate
the dependence of field fluctuations on $k$ at a fixed temperature in Fig. 4.

\medskip
\goodbreak
\bigskip
\centerline{\bf IV. MASSIVE SCALAR ELECTRODYNAMICS}
\xdef\secsym{4.}\global\meqno = 1
\medskip

The results derived in Section II for the $\lambda\phi^4$ theory
can be readily applied to scalar QED which is defined
by the following lagrangian density:
\eqn\qwlag{\eqalign{{\cal L}_{\rm SQED}=&-{1\over4}F_{\mu\nu}F_{\mu\nu}
-{1\over2\alpha}(\partial_\mu A_{\mu})^2\cr
&+\vert(\partial_\mu+ieA_\mu)\phi(x)\vert^2
+{\mu^2\over 2}\phi(x)^\dagger\phi(x)+{\lambda\over 6}
(\phi(x)^\dagger\phi(x))^2,\cr}}
where
$F_{\mu\nu}=\partial_\mu A_\nu-\partial_\nu A_\mu.$
The complex field, $\phi(x)$, will be parameterized as
$(\phi_1(x)+i\phi_2(x))/{\sqrt 2}$, where $\phi_1$ and $\phi_2$ are real.
In computing the scalar QED blocked potential at finite temperature,
it is desirable
that the blocking scale $k$ be introduced into the theory in a
gauge invariant manner for preserving gauge symmetry.
A sharp cut-off smearing function $\rho_k(p)$ which explicitly violates
such symmetry should not be used. The scheme that contains the scale
$k$ and yet is gauge invariant does exist, and it is similar to
the Pauli-Villars
regularization method \ref\reg. The way how $\Lambda$ and $k$ enter gauge
invariantly as the UV and IR regulator, respectively, can easily be seen
by invoking the Schwinger proper-time formalism, in which the functional
determinant of a general second-order operator $\calh$ is represented by:
\eqn\hkorr{ {\rm ln}\Bigl( {{\cal H}\over {\cal H}_0}\Bigr)
= -\int_0^\infty{ds\over s}\Bigl(e^{-{\cal H}s}
-e^{-{\cal H}_0s}\Bigr) ,}
where the expression is normalized with respect to the corresponding
operator $\calh_0$ evaluated at vanishing field. If we introduce
into the proper-time $s$ integration a smearing function
\eqn\sme{\rho(z)=1-(1+z)e^{-z},}
\hkorr\ then becomes, setting $z=\Lambda^2s$,
\eqn\hko{ -\int_0^\infty{ds\over s}\rho(\Lambda^2s)\Bigl(e^{-{\cal H}s}
-e^{-{\cal H}_0s}\Bigr)
={\rm ln}\Bigl[ {{\cal H}\over {\cal H}_0}\times
{{\calh_0+\Lambda^2}\over {\calh+\Lambda^2}}\Bigr]
-{\Lambda^2(\calh-\calh_0)\over {(\Lambda^2+\calh)(\Lambda^2+\calh_0)}},}
which is the Pauli-Villars regularized version containing a UV cut-off.
The manner in which the propagator is modified can also be seen
by the following relation:
\eqn\hkk{{1\over \calh}
=\int_0^\infty ds~ e^{-{\cal H}s}
\longrightarrow \int_0^\infty ds~\rho(\Lambda^2s)e^{-{\cal H}s}
={1\over\calh}\Bigl(1-{\calh\over {\Lambda^2+\calh}}\Bigr)^2.}
If we now simply choose
\eqn\smea{\rho_k(s,\Lambda)=\rho(\Lambda^2s)-\rho(k^2s),}
then $k$ enters into the theory without breaking gauge symmetry.
Therefore, when evaluating one loop contribution
of the blocked potential, one simply makes the substitution:
\eqn\bpot{\eqalign{&\int_k^{\Lambda}\ddp {\rm ln}
\Bigl({{p^2+V''}\over {p^2+V''_0}}\Bigr) \longrightarrow
-\int_0^{\infty}\ddp\int_0^\infty{ds\over s}
\rho_k(s,\Lambda) e^{-p^2s}\Bigl(e^{-V''s}-e^{-V''_0s}\Bigr) \cr
&
=\int\ddp\Biggl\{ {{k^2(V''-V''_0)}\over {(k^2+p^2+V'')(k^2+p^2+V''_0)}}
-{{\Lambda^2(V''-V''_0)}\over {(\Lambda^2+p^2+V'')(\Lambda^2+p^2+V''_0)}} \cr
&\qquad\qquad
+{\rm ln}\Bigl[ {{k^2+p^2+V''}\over {k^2+p^2+V''_0}}
\times {{\Lambda^2+p^2+V''_0}\over {\Lambda^2+p^2+V''}}\Bigr]\Biggr\} \cr
&
={1\over 32\pi^2}\Biggl\{(\Lambda^2-k^2)V''
-(V'')^2{\rm ln}\Bigl({{\Lambda^2+V''}\over k^2+V''}\Bigr)
+\Lambda^4{\rm ln}\Bigl({{\Lambda^2+V''}\over {\Lambda^2+V''_0}}\Bigr)
-k^4{\rm ln}\Bigl({{k^2+V''}\over {k^2+V''_0}}\Bigr)\Biggr\}.}}

Incidentally, both gauge-invariant (proper-time) and gauge non-invariant
(sharp momentum regularized) expressions
give the same result, up to some $\phi$-independent constants.
The two blocking procedures differ in the sense that one classifies
the modes according to their length scale, $k^{-1}$ and the other according
to their weight in the "partition function", $s$.
In the ultraviolet regime where the eigenvalue spectra of the fluctuation
operator is dominated by the kinetic energy term,
blocking transformation introduced
in the preceding Sections and the present construction yield the same
result. However, as
one approaches the infrared regime the contribution from the potential
term becomes more important in determining the
eigenvalues of the fluctuation operator. In this limit, the
length scale $k^{-1}$ would only be characteristic of
a smooth cut-off procedure
in space-time. As long as the underlying low-energy excitations of the system
are extended, the two cut-off methods can be comparable. However, for
a system having localized low-energy states, the functional similarity
exhibited in the two
different blocking approaches is slightly misleading.

In computing the finite-temperature blocked potential it is convenient to
work with Landau gauge $\alpha=0$. With $\Phi^a=\Phi\delta^{a,1}$ and by
performing blocking for the scalar field only
one has the following result:
\eqn\ubbe{\eqalign{
U_{\beta,k}(\Phi) &={\mu_R^2 \over 2}\Phi^2
\Bigl(1-{\lambda_R\over 48\pi^2} \Bigr)
+{\lambda_R \over 4!}\Phi^4\Bigl(1-{5\lambda_R\over 32\pi^2}\Bigr)\cr
&
+{1 \over 32\pi^2}\Biggl\{-k\Bigl(2k^2+\mu_R^2+{\lambda_R\over 2}\Phi^2\Bigr)
\Bigl( {{k^2+\mu_R^2+{\lambda_R\over 2}\Phi^2}}\Bigr)^{1/2} \cr
&
-k\Bigl(2k^2+\mu_R^2+{\lambda_R \over 6}\Phi^2\Bigr)\Bigl( {
k^2+\mu_R^2+{\lambda_R \over 6}\Phi^2}\Bigr)^{1/2} \cr
&
+\Bigl(\mu_R^2+{\lambda_R \over 2}\Phi^2\Bigr)^2 {\rm ln} \Bigl[ {{k + {\sqrt
{k^2+ \mu_R^2+\lambda_R\Phi^2/2}}} \over \mu_R} \Bigr] \cr
&
+\Bigl(\mu_R^2+{\lambda_R \over 6}\Phi^2\Bigr)^2 {\rm ln}\Bigl[{{k+{\sqrt
{k^2+ \mu_R^2+\lambda_R\Phi^2/6}}} \over \mu_R} \Bigr] \Biggr\} \cr
&
+ { 1 \over {2\pi^2\beta}} \Biggl\{  \int_k^{\infty}
dpp^2{\rm ln}\Bigl[1-e^{-{\beta\sqrt{p^2+\mu_R^2
+\lambda_R\Phi^2/2}}}\Bigr]\cr
&
+\int_k^{\infty} dpp^2{\rm ln}\Bigl[1-e^{-{\beta\sqrt{p^2+\mu_R^2
+\lambda_R\Phi^2/6}}}\Bigr]\cr
&
+3\int_0^{\infty} dp p^2 {\rm ln} \Bigl[1-e^{-{\beta\sqrt
{p^2+e^2\Phi^2}}} \Bigr]  \Biggr\},}}
where the renormalized coupling constant $\lambda_R$ is again
defined by an off-shell description \sig.
In the regime where $k^2,T^2 \gg \mu_R^2+\lambda_R\Phi^4/2$
and $e^2\Phi^2,$
\eqn\ugw{U_{\beta,k}(\Phi) \approx{1\over2}\mu_R^2(\beta,k)\Phi^2
+{\lambda_R \over 4!}\Phi^4 +
{3e^4\over 64\pi^2}\Phi^4\Bigl[{\rm ln}\Bigl({\Phi^2\over\mu_R^2}\Bigr) -
{ 25 \over 6} \Bigr] ,}
where
\eqn\tskm{\mu_R^2(\beta,k)= \mu_R^2 - {\lambda_R \over {12\pi^2}}k^2
+{1 \over 6\beta^2} \Biggl[ {2\lambda_R\over \pi^2}\sum_{n=1}^{\infty}
e^{-n\beta k}\Bigl({1\over n^2}+{{\beta k}\over n}\Bigr)
+{3\over 2} e^2 \Biggr]. }
The minimum of $U_{\beta,k}(\Phi)$ is again located by
\eqn\kor{ 0 = \mu_R^2(\beta,k) + \Phi^2\Bigl( {\lambda_R \over 6} -
{11e^4 \over 16\pi^2} \Bigr) + {3e^4\Phi^2 \over 16\pi^2}{\rm ln}
\Bigl({\Phi^2 \over \mu_R^2}\Bigr). }
We first note that for $U_{1/{\beta}=k=0}(\Phi)$ to have non-trivial
minimum, the condition
\eqn\kgp{ \lambda_R < {9e^4 \over 8\pi^2} \Biggl\{ {8 \over 3}
-  {\rm ln} \Bigl( {16\pi^2 \over 3e^4}
\Bigr) \Biggr\} }
must be satisfied.
For small $e$, the expression above leads to a negative
quartic coupling $\lambda_R$ for the theory, which is in contradiction
with our original assumption.
Therefore, as long as $\lambda_R >0$, $U_{1/{\beta}=k=0}(\Phi)$
always has unique minimum at $\Phi=0$ for the theory
in the symmetric phase, with $\mu_R^2 >0$.
Again, we plot the $k$ dependence of $U_{\beta,k}(\Phi)$ for a given set
of parameters in Fig. 5.

\medskip
\goodbreak
\bigskip
\centerline{\bf V. DYNAMICAL MASS GENERATION }
\xdef\secsym{5.}\global\meqno = 1
\medskip
\nobreak

We explore in this Section the mechanism of mass parameter
generation according to
the results obtained before. As stated in the Introduction,
the novelty of this construction is that it is in effect
even when the expectation value of the scalar field is vanishing.
The foundation of this mechanism is based on the following three
arguments:
(1) within a sufficiently small region $\calr$ such that the linear
dimension of $\calr$ is smaller than $\varsigma\sim k_{cr}^{-1}$,
the correlation length, the scalar field fluctuates
around different locally non-vanishing values even in the symmetrical phase
with an overall zero field expectation value;
(2) what is relevant from the point of view of the mass generation
is the behavior of the scalar field within
the region where the particle propagates, and not the whole space-time;
and (3) at sufficiently
high temperature, $\calr$ is so short in the time direction
that the local field values are correlated inside. As we shall see below,
if this effective propagation region is smaller
than the domain size, then
within a certain range of coupling constants
it is possible to obtain the non-trivial solution for mass parameter in
a self-consistent manner.

Consider now a fermionic test
particle described by the field $\psi(x)$ and the lagrangian
\eqn\testl{{\cal L}=\bar\psi(x)(\thru\partial+{\cal U}(x)
+G\phi(x))\psi(x).}
We choose a fermion field here but our qualitative remarks hold for
scalar as well as vector fields.
Suppose the static external potential,
$\cal U$  is strong enough to create a bound state
of size $L$ consisting of a very heavy anti-fermion and the massless test
particle. The bound state wave function will then be localized in a
space-time ``tube'' $\cal R$ which is infinitely long in the time direction
and has a width $L$ in the spatial direction.
The fermion, due to the binding effect,
is constrained to stay in the vicinity of
the heavy one. Thus, the region of space-time
traced by the fermion is practically a tube consisting of
points with $\s x^2<1/k^2$, where $1/k$ is the tube radius characterizing
the size of the bound state.

What determines the mass parameter for the
fermion in this bound state ? It is perhaps much easier to answer this
question by inquiring what is unimportant
from the fermion's point of view. Obviously, how the scalar field behaves
outside the tube is irrelevant since the fermion
is never there. The mass parameter
should then be determined solely by the dynamics of the scalar field,
$\phi(x)$, $within$ the tube.

The binding
mechanism alone is sufficient to generate a non-vanishing
mass parameter which otherwise would have been prohibited from the
symmetry argument of the lagrangian. A renowned example to substantiate
the effect of binding on mass generation is the chiral symmetry breaking
and the generation of valence quark mass in the bag model. In this model,
it is just the boundary condition
on the quark wave functions which leads to the mixing of
small and large spinor components and generation of
mass parameter by "binding" the wave functions to fit inside the bag.

Since it is the fluctuations
of the scalar field $\phi(x)$ within the tube that determine the
mass parameter of the bound test particle through the Yukawa coupling in
\testl, one simply concentrates on the interior of the
tube and isolates the "most important" mean field values for $\phi(x)$
from determining the maxima of the distribution function:
\eqn\distrim{\varrho_M(\Phi)=<\delta(\Phi-\phi_{\cal R})>
=\int D[\phi]e^{iS_M[\phi]}\delta(\Phi-\phi_{\cal R}),}
where $S_M[\phi]$ is the Minkowski space-time action and the blocked
variable $\phi_{\cal R}$ is constructed by \blvi\ in Minkowski space-time
as well. The relation between the distribution function \distrim\ and
the local potential in the blocked action, $U^{(0)}_{\cal R}(\Phi)$
is given by \locapr.
The surface to volume ratio of the tube $\cal R$ stays finite
as the time extent of the tube tends to infinity. Thus the coupling between
``neighboring'' tubes is strong and the approximation given in \locapr\
becomes invalid.

However, we have good reason to trust \locapr\ at high temperature.
This is because then the fluctuations are strongly damped
and suppress the contributions of the derivative terms
in the blocked action. A more formal way to see this is to note that
the time extent of the Euclidean system is small at high temperature
and thus \locapr\ is justified. Since the distribution of the static modes
agree in the Minkowski and the Euclidean space-time, \locapr\
holds for both cases as well. The results presented above indicates
that there is a
non-zero mass parameter generated in this manner for small enough
localized states either in the Yukawa-model or in scalar QED.

Fundamentally, there are differences in the way
bosons and fermions acquire mass; namely, the former couples
quadratically to the scalar field, while the latter couples only linearly.
For the bosons, although  the two most probable Higgs
field averages $\pm\Phi_{\beta,k}$, differ in sign, the mass
squared generated from either contribution nevertheless agree; however,
for fermions, the two background field values lead
to opposite signs for the generated mass. We argue that the fermionic Green's
functions must retain the chirality-even sector only. In fact, since chiral
transformation changes the sign of the mass term due to
the presence of $\gamma_5$, only the contributions with even chirality
will survive the averaging over $\pm\Phi_{\beta,k}$.
In that case, we expect the single-particle fermionic propagator
to be different from the usual one. The propagator for the bound state of
even number of fermions will remain the same, however.

The above consideration for a bound state system
is not particularly illuminating. The truly challenging and important issue
that remains, is how to implement such mechanism to point-like particles,
in which the presence of a non-zero mass parameter is excluded at the tree
level by the gauge symmetry of the lagrangian. Let us consider now a
point-like, free test particle described by the lagrangian:
\eqn\testsf{\eqalign{{\cal L}&=
\bar\psi(x)(\thru\partial+G\phi(x))\bar\psi(x)\psi(x)\cr
&=\bar\psi(x)(\thru\partial+m)\psi(x)
+(G\phi(x)-m)\bar\psi(x)\psi(x)\cr
&=\bar\psi K_0\psi+\bar\psi\delta K\psi,\cr}}
where the ``variational'' mass parameter, $m$,
is to be chosen in a $\phi(x)$-dependent
manner such that the perturbation expansion in $\delta K$ for the
propagator,
\eqn\pertprt{<G>=<{1\over K_0+\delta K}>=\int D[\phi]e^{-S[\phi]}
{1\over K_0[\phi]+\delta K[\phi]},}
is optimized.

In order to determine the form of $m[\phi]$, we must first find
approximatively the space-time region
where the test particle propagates. In the zeroth order of $\delta K$,
we follow the proper-time parametrization due to Schwinger and write
\eqn\propsch{G(x,y)=<x|{m-\thru\partial\over m^2-\partial^2}|y>=
(m-\thru\partial)\int_0^\infty ds<x|e^{-s(m^2-\partial^2)}|y>.}
The matrix element in the second equation can be obtained in the
path integral representation. In fact, it is the Euclidean time evolution
operator matrix element corresponding to the "hamiltonian"
\eqn\hamsch{H=m^2-\partial^2,}
which acts on the "wave functions" defined over the four dimensional
space-time. The repetition of the standard steps lead to the
path integral formulae for our free 4+1 dimensional problem,
\eqn\propt{
G(x,y)=(m-\thru\partial)\int_0^\infty ds\int_{z(0)=x}^{z(s)=y}
D[z^\mu(\tau)]e^{-sm^2-{1\over4}\int_0^sd\tau\dot z^2(\tau)}.}
Here the functional integration is performed over trajectories in space-time
which connect the points $x$ and $y$ as the proper time, $\tau$, moves from
0 to $s$. The functional integral is Gaussian and can be carried out
to give:
\eqn\propts{G(x,y)={1\over16\pi^2}(m-\thru\partial)\int_0^\infty{ds\over s^2}
e^{-sm^2-{(x-y)^2\over4s}},}
This is the immediate consequence of the relation
\eqn\nonrtd{<x|e^{s{\partial^2\over\partial x^2}}|y>=
{1\over\sqrt{4\pi s}}e^{-{(x-y)^2\over4s}},}
for the one dimensional propagator in quantum mechanics.

The trajectories in the functional
integration are restricted to be within a finite-width world tube, the
shape of which would then determine the distribution of the field
$\phi(x)$ experienced by the test particle. Having obtained the world
tube that corresponds to the test particle propagator, it is
sufficient to
consider only the behavior of the scalar field within this tube.
The non-Poincare invariant aspect of this reasoning poses no difficulty
since the location and shape of the tube introduced here are also
dependent on the locality of creation and annihilation
of the test particle.

In evaluating $G(x,y)$,
the propagator of a free, unbound particle at large space-time
separation, $x-y\to\infty$, there are in principle, two
relevant length scales which appear in the distribution of the particle
location in the hyperplane perpendicular to $x-y$.
The first one is the width of the world tube,
$\ell$, which gives account of the region of space-time
traversed by the particle. The other
length scale, $S$, is the characteristic size of the particle itself.
These two length scales are independent. In fact, $S$ depends on the
dynamics of the constituents of the particle. This is obvious for
a composite particle. For a
point particle we have the unavoidable particle-anti particle cloud
which generates the finite extent $S$. It will be sufficient to
determine $\ell$ for our purposes. Since $m$ is the only scale parameter
of the problem we must have
\eqn\dimant{\ell~\approx {c\over m}.}

We are now in the position to determine the optimal choice of $m$.
To minimize the effect of $\delta K$ we obviously have
to choose
\eqn\optch{m=\pm G\Phi_{\beta,k},}
where $\Phi_{\beta,k}$ is the most probable average field
for the world tube of the unbound test particle. The sign of $m$ follows
that of the actual average field in \pertprt.
This relation leaves the width of the world tube, $\ell=k^{-1}$, undetermined.
The self-consistent choice of $k$ is such that
in the corresponding world tube, the most probable scalar field average,
$\Phi_{\beta,k}$, reproduces the same mass parameter, i.e.
\eqn\consf{G\Phi_{\beta,{k=m/c}}=m.}
Note that this relation holds in the symmetry broken phase as well.

We can now turn to the models discussed in the previous Sections.
For the Yukawa theory considered in Section III,
the self-consistent equations for the fermion mass parameter are
\eqn\yede{\cases{\eqalign{0&= \mu_R^2(\beta,k)+\Phi_{\beta,k}^2\Bigl(
{\lambda_R \over 6} + {11g^4 \over 12\pi^2}\Bigr)
-{g^4\Phi_{\beta,k}^2 \over 4\pi^2}{\rm ln}\Bigl({\Phi_{\beta,k}^2\over
\mu_R^2}\Bigr) \cr
M_{\psi}&= g\Phi_{\beta,k}=c_1k ,\cr}}}
which is reduced to
\eqn\mtpa{\eqalign{ 0&=\mu_R^2-{\lambda_R\mpsi^2\over {48\pi^2g^2c_1^2}}
\Bigl\{ 3g^2\Bigl[1+{4g^2c_1^2\over\lambda_R}{\rm ln}\Bigl({\mpsi^2\over
{g^2\mu_R^2}}\Bigr)\Bigr]-8c_1^2\pi^2\Bigl(1+{11g^4\over {2\lambda_R\pi^2}}
\Bigr)\Bigr\} \cr
&
-{g^2T^2\over 3}\Bigl[1-{3\lambda_R\over 4g^2\pi^2}\sum_{n=1}^{\infty}
e^{-n\mpsi\over Tc_1}\Bigl({1\over n^2}+{\mpsi\over nTc_1}\Bigr)\Bigr],}}
upon use of \exis. The mass parameter $M_{\psi}(T)$ is plotted in Fig. 6
for $\lambda_R=0.4$, $c=0.01$ and $g=0.2$ and $0.15$. Note that the validity
of $M_{\psi}$ for low $T$ may become questionable since the formalism of
mass parameter is expected to break down. Qualitatively, our result is
in agreement with \ref\suzhou\ in which the author obtained a larger
Euclidean mass parameter with increasing Matsubara frequencies,
although our conclusion is based
on a different, mean-field argument. The decrease of the mass parameter
at lower temperature may not be significant since \locapr\ is no longer
appropriate.

For the massive scalar QED, we shall consider its impact of on the mass
parameter of another scalar particle $\varphi$. The lagrangian of the
entire system is written as
\eqn\tlag{ {\cal L}={1\over 2}(\partial_{\mu}\varphi)^2
+{1\over 2}G\phi^2\varphi^2+{\cal L}_{\rm SQED},}
where ${\cal L}_{\rm SQED}$ is given by \qwlag. Even though the background
Higgs field may have vanishing vacuum expectation value, $<\phi(x)>=0$,
$<\phi^2(x)>$ needs not necessarily be zero, as we have shown before.
The leading mean-field contribution to the mass parameter of
$\varphi$ would then be $G\phi^2$, where $G$ is the interaction strength.
In this case, the self consistent equations for the scalar particle in
scalar QED would be
\eqn\qede{\cases{\eqalign{ 0 &=\mu_R^2(\beta,k)+\Phi_{\beta,k}^2\Bigl(
{\lambda_R \over 6} - {11e^4 \over 16\pi^2} \Bigr)
+{3e^4\Phi_{\beta,k}^2 \over 16\pi^2}{\rm ln}\Bigl({\Phi_{\beta,k}^2
\over \mu_R^2}\Bigr) \cr
M_{\varphi}&= G\Phi_{\beta,k}=c_2k, \cr}}}
or
\eqn\mtp{\eqalign{ 0&=\mu_R^2-{\lambda_R\mvar^2\over {48\pi^2G^2c_2^2}}
\Bigl\{ 4G^2\Bigl[1-{9e^4c_2^2\over 4\lambda_R}{\rm ln}
\Bigl({\mvar^2\over{G^2\mu_R^2}}\Bigr)\Bigr]-8c_2^2\pi^2\Bigl(1-{33e^4
\over {8\lambda_R\pi^2}}\Bigr)\Bigr\} \cr
&
+{e^2T^2\over 4}\Bigl[1+{4\lambda_R\over 3e^2\pi^2}\sum_{n=1}^{\infty}
e^{-n\mvar\over Tc_2}\Bigl({1\over n^2}+{\mvar\over nTc_2}\Bigr)\Bigr].}}
Once more, this non-linear equation can only be solved for certain values of
coupling constants, and in general they may not be unique.
In Fig. 7, we depict the $T$ dependence of the mass parameter for
$c_2=0.05$ and $c_2=0.1$. There, we observe a very slow increase in
$\mvar$ for low $T$, followed by a rapid growth. The interpretation is
qualitatively the same as that of Yukawa model.

In summary, we find that at high temperature where the static modes
dominate the Euclidean path integral, a test particle which propagates
and interacts with the background Higgs field can acquire a non-vanishing
mass parameter in the mean-field approximation if its path of propagation
is a narrow tube within which the scalar Higgs field has non-vanishing
most probable values. If it scans through a large region of space-time,
then the average Higgs field would certainly vanish. However,
noting that how the
particle propagates is directly influenced by this mass parameter, we
solve the coupled self-consistent equations and conclude that the mass
parameter for the test particle may be non-vanishing for a chosen set of
coupling constants.

\medskip
\goodbreak
\bigskip
\centerline{\bf V. SUMMARY}
\xdef\secsym{5.}\global\meqno = 1
\medskip
We have shown that in a certain temperature range, the finite-temperature
blocked potential $U_{\beta,k}(\Phi)$ develops minima at
non-vanishing field values for short enough length scale,
$k^{-1}<k^{-1}_{cr}$, even in the symmetrical phase, and that
the appearance of such degenerate minima differs from the phenomenon
of spontaneous symmetry breaking.
Although the field average computed in the finite region
fluctuates around two symmetrical non-zero values, this
has nothing to do with the behavior of the extreme infrared mode,
$\phi_{k=0}={1\over\Omega}\int d^4x\phi(x)$,
$\Omega=\int d^4x$. All we see is a distinct scale
dependence of the fluctuations which suggests
that the typical field configurations of the $symmetrical$
phase contain domains, $\phi(x)\sim\pm\Phi_{\beta,k_{cr}}$,
of the size $\varsigma\sim k^{-1}_{cr}$.
This is because the blocked variables fluctuate around zero or
$\pm\Phi_{\beta,k}$ for $\varsigma>k_{cr}^{-1}$ or $\varsigma<k_{cr}^{-1}$,
respectively, in the presence of the domains.

Various interesting directions
may be pursued starting with this result. One is that the Landau-Ginsburg
theory may exhibit domain structure even in the symmetrical phase.
That is, the fields may fluctuate around non-zero values even in the absence
of long-range order.
For example, although an Ising system becomes paramagnetic
with average magnetization ${\cal M}=0$ at $T > T_c,$ where
$T_c$ is the Curie temperature, there may exist local, ferromagnetic
domains with  ${\cal M}=\pm 1.$
The size of a typical domain should be $O(k^{-1}_{cr})$, a decreasing function
of the temperature.

The possibility of mass parameter generation without symmetry
breaking was explored in this paper by
considering a Higgs field coupled to massless fermions via Yukawa coupling,
or to another scalar particle in the scalar QED background.
Within a flux tube of characteristic width $k^{-1}$ along the world line,
the fermions
feel the fluctuations of the Higgs field within a ``radius'' $\sim k^{-1}$.
If this is smaller than $k_{cr}^{-1}$, then the Higgs field
may generate mass for fermion even if the theory is in the symmetrical
phase.
It was found that the temperature generates a thermal mass and constraints
the blocked variable to be within distance $O(1/T)$ from zero.
This result is in agreement with the expectation that the domain size should
decrease as the temperature is increased.
Therefore, when the effects of the heat bath are included, we need a
larger scale energy in order to observe mass generation.
Consequently, the production of small mass particles is suppressed.

It seems that this alternative mechanism is no more realistic
than the one based on spontaneous symmetry breaking, for it too,
involves a scalar ``Higgs'' field, albeit in a manner which breaks no
symmetry. Despite the fact that our presentation
here offers only a rough, mean-field
solution to mass parameter generation, we believe that this novel approach
for the old problem is worth communicating even at its rudimentary
level, in order to stimulate further thoughts and explorations
along this direction. The fact that the
self-consistent conditions given in Section V for
deducing mass parameters for unbound particles
could only be satisfied by choosing small $c$ may also
be undesirable for investigating the feed-back to the scalar
field from the test particle.
Nevertheless, it can be taken into account with
strong coupling expansion and we plan to return to this problem in a
separate publication.

We also comment that phenomena similar to our discussion of mass generation
have also been considered before for two-dimensional theories
in which dynamical breakdown of continuous symmetries is not possible due
to the severe infrared
divergences \ref\mermwag. Mass generation in the $SU(N)$ Thirring
\ref\witt\ and Yukawa \ref\jers\ models in $d=2$ has been attained,
with long-range correlations in the absence of any symmetry breaking effect.

We conclude that the temperature dependence is a useful tool to trace
back the role the field fluctuations play in the theory. It allows us to
vary domain size and mass parameters in a distinctive manner. It would be
interesting to observe such effects experimentally in some easily accessible
systems.

\medskip
\goodbreak
\bigskip
\centerline{\bf ACKNOWLEDGEMENT}
\medskip
\nobreak
S.-B. L. is grateful to C. Gong for fruitful discussions.
\bigskip
\centerline{\bf FIGURE CAPTIONS}
\medskip
\nobreak
\par\hang\noindent
Fig. 1. Momentum space representation of $O(3)$ and $O(4)$ symmetric
smearing function.
\medskip
\par\hang\noindent Fig. 2. Dependence of
$U_{\beta,k}(\Phi)$ on $k$ for $\lambda_R=0.1$ and $T=20$, in units of $\mu_R$.
\medskip
\par\hang\noindent Fig. 3. Dependence of
$U_{\beta,k}(\Phi)$ on $T$ for $\lambda_R=0.1$ and $k=50$ in unit of $\mu_R$.
Fluctuations around non-trivial minima will disappear for $T \ge T_k$.
\medskip
\par\hang\noindent
Fig. 4. Dependence of $U_{\beta,k}(\Phi)$ on $k$ for Yukawa model
with $\lambda_R=0.4$, $g=0.2$ and $T=15$, in units of $\mu_R$.
\medskip
\par\hang\noindent
Fig. 5. Dependence of $U_{\beta,k}(\Phi)$ on $k$ for Massive scalar QED
with $\lambda_R=0.4$, $e=0.02$ and $T=20$, in units of $\mu_R$.
\medskip
\par\hang\noindent
Fig. 6. Temperature dependence of the self-consistent mass parameter
$\mpsi$ in Yukawa model.
\medskip
\par\hang\noindent
Fig. 7. Temperature dependence of the self-consistent mass parameter
$\mvar$ in massive scalar QED.
\medskip

\goodbreak
\bigskip
\centerline{\bf REFERENCES}
\medskip
\nobreak
\par\hang\noindent{\wils} K. Wilson and J. Kogut, {\it Phys. Rep.}
{\bf 12C} (1975) 75.
\medskip
\par\hang\noindent{\lp} S.-B. Liao and J. Polonyi,
{\it Ann. Phys.} {\bf 222} (1993) 122.
\medskip
\par\hang\noindent{\wet} see for example, C. Wetterich,
{\it Nucl. Phys.} {\bf B352} (1991) 529;
M. Reuter and C. Wetterich, preprint DESY 92-037; N. Tetradis and C. Wetterich,
HD-THEP-93-36 and references therein.
\medskip
\par\hang\noindent{\dimred} D. Gross, R. Pisarski and A. Yaffe,
{\it Rev. Mod. Phys.} Vol. {\bf 53} {No. 1}, (1981) 43.
\medskip
\par\hang\noindent{\tsypin} We thank M. Tsypin for pointing this out for us.
\medskip
\par\hang\noindent{\col} S. Coleman and E. Weinberg, {\it Phys. Rev.}
{\bf D7} (1973) 1888.
\medskip
\par\hang\noindent{\arfken} G. Arfken, {\it Mathematical Methods for
Physicists}, p. 338, (Academic Press, New York, 1985).
\medskip
\par\hang\noindent{\wein} S. Weinberg, {\it Phys. Rev.}
{\bf D9} (1974) 3357.
\medskip
\par\hang\noindent{\pen} P. Fendley, {\it Phys. Lett.}
{\bf B196} (1987) 175; P. Elmfors, {\it Z. Phys.} {\bf C56} (1992) 601
and references therein.
\medskip
\par\hang\noindent{\funa} K. Funakubo and M. Sakamoto, {\it Phys. Lett.}
{\bf B186} (1987) 205.
\medskip
\par\hang\noindent{\jack} L. Dolan and R. Jackiw, {\it Phys. Rev.}
{\bf D9} (1974) 3320.
\medskip
\par\hang\noindent{\lin} A. Linde, {\it Rep. Prog. Phys.} Vol. {\bf 42}
(1979) 379;
D. A. Kirzhnits and A. D. Linde, {\it Phys. Lett.} {\bf 42B} (1972) 471.
\medskip
\par\hang\noindent{\reg} M. Oleszczuk, ``A symmetries Preserving Cut-Off
Regularization''; S.-B. Liao and J. Polonyi, Duke-TH-94-65.
\medskip
\par\hang\noindent{\suzhou} S. Huang, {\it Phys. Rev.} {\bf D47} (1993) 653.
\medskip
\par\hang\noindent{\mermwag} N. D. Mermin and H. Wagner, {\it Phys. Rev. Lett.}
{\bf 17} (1966) 1133; S. Coleman, {\it Comm. Math. Phys.}
{\bf 31} (1973) 259.
\medskip
\par\hang\noindent{\witt} E. Witten, {\it Nucl. Phys.} {\bf B145} (1978) 110.
\medskip
\par\hang\noindent{\jers} A. K. De, E. Focht, W. Franzki, J. Jersak and
M. A. Stephanov, `Study of the asymptotic freedom of 2d Yukawa
models on the lattice', J\"ulich preprint, HLRZ 93-15.
\medskip

\goodbreak
\bigskip
\centerline{\bf APPENDIX A}
\xdef\secsym{\rm A.}\global\meqno = 1
\medskip
\nobreak
In this Appendix, we give an estimate of the
width, $\ell$, of the tube where the
test particle with mass $M$ propagates by computing
$<<z_{tr}^2>>_{tr}$. Here $z_{tr}$ stands for the transverse part of
$z-x$:
\eqn\ztr{z_{tr}=z-x-(y-x){(z-x)\cdot(y-x)\over(y-x)^2}.}
The double bracket,
$<<\cdots>>_{tr}$, denotes the normalized
averaging over the trajectories in the
functional integration \propt\ of the average along the trajectories.
The latter, the averaging along the trajectories can be obtained by
splitting the proper time, $s$, into two subsequent parts,
$s=t+u$. The point, $z$, on the trajectory is then taken at the proper time
$t$ and the values of $t$ and $u$ are integrated over,
\eqn\ztrao{\eqalign{<<z^2_{tr}>>_{tr}&=
{\cal N}^{-1}\int_0^\infty{ds\over s}\int_0^\infty dt\int_0^\infty
du\delta(s-t-u)\int D[z]z^2(t)_{tr}
e^{-m^2(t+u)-{1\over4}\int_0^sd\tau \dot z^2(\tau)}\cr
&={\cal N}^{-1}\int_0^\infty dt\int_0^\infty du{1\over t+u}\int D[z]z^2(t)_{tr}
e^{-m^2(t+u)-{1\over4}\int_0^{t+u}d\tau \dot z^2(\tau)},\cr}}
where ${\cal N}^{-1}$ is the normalization constant,
\eqn\ztrano{{\cal N}=
\int_0^\infty ds\int D[z]
e^{-m^2s-{1\over4}\int_0^sd\tau \dot z^2(\tau)}.}

The functional integration is now written as the integral
over trajectories corresponding to the proper time intervals $0<\tau<t$
and $t<\tau<t+u$ and connecting the space-time points $x,z$ and $z,y$,
respectively when $z$ is integrated over, as well. The functional
integration can be carried out yielding
\eqn\ztrat{<<z^2_{tr}>>_{tr}={1\over(4\pi)^4}
\int{dt~du\over(t+u)t^2u^2}\int d^4z z^2_{tr}
e^{-m^2(t+u)-{(z-x)^2\over4t}-{(y-z)^2\over4u}}.}
The $t$ and $u$ integrals are evaluated in the saddle point approximation.
By retaining the exponent in the selection of the saddle point we have,
$t_0=|z-x|/2m$ and $u_0=|y-z|/2m$, which give
\eqn\sadztr{<<z^2_{tr}>>_{tr}=m^2{{\cal N}^{-1}\over16\pi^3}
\int d^4z z^2_{tr}{\sqrt{|z-x||y-z|}\over(|z-x|+|y-z|)(z-x)^2(y-z)^2}
e^{-m(|z-x|+|y-z|)}.}
Upon rescaling the variables, $y\to my$, $z\to mz$, we obtain, for
$x=0$,
\eqn\proptw{<<z^2_{tr}>>_{tr}={1\over m^2}{{\cal N}^{-1}\over16\pi^3}
\int d^4z\bigl(z^2-(n\cdot z)^2\bigr)
{\sqrt{|z-x||y-z|}\over(|z-x|+|y-z|)(z-x)^2(y-z)^2}e^{-|z|-|y-z|},}
where $n={y\over|y|}$. The normalization can be obtained in a similar
manner,
\eqn\propn{{\cal N}={1\over16\pi^3}
\int d^4z{\sqrt{|z-x||y-z|}\over(|z-x|+|y-z|)(z-x)^2(y-z)^2}
e^{-|z|-|y-z|},}
yielding
\eqn\normav{<<z^2_{tr}>>_{tr}={1\over m^2}{\rm lim}_{y\to\infty}
{\int d^4z\bigl(z^2-(n\cdot z)^2\bigr)
{\sqrt{|z-x||y-z|}\over(|z-x|+|y-z|)(z-x)^2(y-z)^2}
e^{-|z|-|y-z|}\over\int d^4z{\sqrt{|z-x||y-z|}\over(|z-x|+|y-z|)(z-x)^2(y-z)^2}
e^{-|z|-|y-z|}}.}
Note that the ratio on the right hand side is convergent. Thus we have
\eqn\diman{<<|z_{tr}|>>_{tr}~\approx {c\over m}.}

\bigskip
\centerline{\bf APPENDIX B}
\xdef\secsym{\rm B.}\global\meqno = 1
\medskip
\nobreak
\medskip

For completeness, we turn to massless scalar QED with $\mu_R^2 =0,$
and consider the following cases.

\vskip 12pt
\noindent (1) $k^2=1/{\beta}^2=0:$

The potential here reads as
\eqn\fuj{U_{1/{\beta}=k=0}(\Phi) =
{\lambda_R \over 4!}\Phi^4 + {3e^4\Phi^4 \over {64 \pi^2}}
\Bigl[{\rm ln}\Bigl({\Phi^2\over M^2}\Bigr)-{25\over 6} \Bigr],}
where we have assumed a small value of $\lambda_R \sim O(e^4).$
As shown in \col, the symmetry of the system is spontaneously broken
due to the radiative corrections, with minimum of the potential
being $M$. Dimensional transmutation also takes place with the scalar
field coupling related to the gauge field coupling
strength by
\eqn\edl{ \lambda_R = {33e^4 \over 8\pi^2},}
which leads to the familiar result:
\eqn\kop{ U_{ {1 /{\beta}=k=0}}(\Phi) =
{3e^4\Phi^4 \over {64 \pi^2}}
\Bigl[{\rm ln}\Bigl( { \Phi^2 \over M^2}\Bigr)-{1\over 2} \Bigr].}
The parameter space spanned by the the potential is now $(e, M)$,
instead of the original $(e,\lambda_R)$.
\vskip 12pt
\noindent (2) $k^2 =0, 1/{\beta}^2 \ne 0:$

If we consider only the temperature effects, then for
${1/{\beta^2}} \gg \mu_R^2+\lambda_R\Phi^4/2$ and $e^2\Phi^2,$
\eqn\fde{ U_{\beta,k=0}(\Phi) =
{ {e^2 } \over 8\beta^2}\Phi^2
+{\lambda_R \over 4!}\Phi^4 + {3e^4\Phi^4 \over {64 \pi^2}}
\Bigl[{\rm ln}\Bigl({\Phi^2\over M^2}\Bigr)-{25\over 6}\Bigr] .}
Using \edl, the expression above is reduced to
\eqn\poi{U_{\beta,k=0}(\Phi) =  {{e^2} \over 8\beta^2}\Phi^2 +
{3e^4\Phi^4 \over {64 \pi^2}}
\Bigl[{\rm ln}\Bigl({\Phi^2\over M^2}\Bigr)-{1\over 2} \Bigr].}
The symmetry for this theory is broken initially due to
radiative correction. However, we anticipate its
restoration at a temperature above $1/{\beta_c},$ which may be found by
solving the equation
\eqn\ter{ {1 \over \beta_c^2} = {\rm min} \Bigl\{ {{3e^2\Phi^2}\over 4\pi^2}
{\rm ln}\Bigl({M^2\over\Phi^2}\Bigr)\Bigr \}.}
With ${\rm ln}\bigl(M^2/{\Phi^2}\bigr)=1$, we obtain
\eqn\tc{ {\rm ln}\Bigl({3e^2M^2\beta_c^2 \over 4\pi^2}\Bigr)=1 }
for $\beta_c$. The pattern of symmetry restoration is first order.
Above the critical temperature, $ U_{\beta,k=0}(\Phi)$
again has unique minimum  at $\Phi=0.$

\vskip 12pt
\noindent (3) $ k^2 \ne 0, 1/{\beta^2}=0:$

As shown in case $(1)$, the symmetry of
the system is spontaneously broken at $k=1/{\beta}=0$, and shall
remain so for non-zero $k$, since, being an internal parameters
characterizing the scale of observation,
it cannot change the phase of the system.
Taking into account the dominant contribution of $k$ in the limit
$k^2 \gg \lambda_R\Phi^2/2$ and $e^2\Phi^2$ gives
\eqn\fuji{U_{ {1/{\beta} }=0,k}(\Phi) =
- { {\lambda_R k^2} \over 24\pi^2}\Phi^2
+{\lambda_R \over 4!}\Phi^4 + {3e^4\Phi^4 \over {64 \pi^2}}
\Bigl[{\rm ln}\Bigl({\Phi^2\over M^2}\Bigr)-{25\over 6} \Bigr],}
from which we obtain
\eqn\fujik{ 0 = {\partial\over {\partial\Phi}} U_{ {1/{\beta}}=0,k}(\Phi)
= \Phi_k \Biggl\{-{{\lambda_R k^2}\over 12\pi^2}+\Phi_k^2\Bigl[
{\lambda_R \over 6}-{11e^4 \over 16\pi^2}\Bigr]+{{3e^4\Phi_k^2}
\over 16\pi^2}{\rm ln}\Bigl({\Phi_k^2\over M^2}\Bigr)\Biggr\}.}
The presence of the $k$-dependent term is to shift the minimum of
$ U_{1/{\beta}=0,k}(\Phi)$ from $M$ to $\Phi_k > M$.
which may be found by solving
\eqn\des{ \Phi_k^2~{\rm ln} \Bigl({{\Phi_k^2} \over M^2}\Bigr)
= { 11 k^2 \over 6\pi^2} .}
The result suggests that in the case of spontaneous symmetry breaking,
depending on the scale, the "effective" vacuum expectation value
$\Phi_k$ may actually be greater than the usual
$M=246$ GeV in the Standard Model.
The smaller the $k^{-1}$, the larger the $\Phi_k$.

\vskip 12pt
\noindent (4) $k^2, 1/{\beta^2} < 1/{\beta_c^2} \ne 0:$

When both the effects due to temperature and scale are considered, we
arrive at
\eqn\cde{ U_{\beta,k}(\Phi) = {1\over2}{\tilde \mu_R^2(\beta,k)} \Phi^2
+ {\lambda_R \over 4!} {\Phi}^4 +
{ 3e^4 \Phi^4 \over {64 \pi^2}}
\Bigl[{\rm ln}\Bigl({\Phi^2\over M^2}\Bigr)-{25\over 6}\Bigr],}
where
\eqn\der{ {\tilde \mu_R^2(\beta,k)} = - {\lambda_R \over {12\pi^2}}k^2
+{1 \over 6\beta^2} \Biggl[ {2\lambda_R\over \pi^2}\sum_{n=1}^{\infty}
e^{-n\beta k}\Bigl({1\over n^2}+{{\beta k}\over n}\Bigr)
+{3\over 2} e^2 \Biggr]. }
As in the massive case, the field fluctuations due to the scale-dependent
term is much suppressed by the high-temperature contribution. To
observe field fluctuations around non-trivial minima, we again must go
to the large $k$ limit.

\vfill
\eject
\end